\documentclass[11pt]{article}
\usepackage{subcaption}
\usepackage{caption}
\usepackage{booktabs}
\usepackage{afterpage}
\usepackage{changepage}
\usepackage{float}
\usepackage{setspace}

\usepackage{amsmath,amsfonts,amsthm,amssymb,latexsym}
\usepackage{bm,bbold}
\usepackage{comment}
\usepackage{graphicx}
\usepackage[margin=1in,letterpaper]{geometry}
\usepackage{indentfirst}
\usepackage{natbib}
\usepackage{soul}
\usepackage{enumitem}

\usepackage[ruled]{algorithm2e}   %
\SetKwInput{KwInput}{Input} 
\SetKwInput{KwOutput}{Output}

\newcommand{\lp}[1]{\left(#1\right)}

\newcommand{\lB}[1]{\left\{#1\right\}}

\newcommand{\lV}[2][]{\left\|#2\right\|_{#1}}

\DeclareMathOperator*{\argmin}{arg\,min}

\newcommand{\indicator}[1]{\mathbb{1}_{\lB{#1}}}

\newcommand{\R}{\mathbb{R}}

\newcommand{\E}[2][]{\mathbb{E}_{#1}\left[#2\right]}

\newcommand{\bmx}{\bm{x}}

\newcommand{\bmS}{\bm{S}}

\newcommand{\bmtheta}{\bm{\theta}}

\newcommand{\bfone}{\mathbf{1}}

\newcount\Comments  %
\usepackage{color}

\usepackage{hyperref}
\usepackage[table]{xcolor}
\hypersetup{
	colorlinks,
	linkcolor={blue!50!black},
	citecolor={blue!50!black},
	urlcolor={blue!80!black}
}
\usepackage{tablefootnote}

\title{
Downside Risk Reduction Using Regime-Switching Signals: \\A Statistical Jump Model Approach\footnote{
This article was previously titled ``Regime-Aware Asset Allocation: a Statistical Jump Model Approach''. 
The authors are grateful for the thoughtful comments from the editor and the anonymous reviewer, which have significantly improved the article.
}
}

\author{
Yizhan Shu\footnote
{Department of Operations Research and Financial Engineering, Princeton University, Princeton, USA} \and
Chenyu Yu\footnotemark[2]  \and
John M. Mulvey\footnotemark[2]  
\,\footnote{
Yizhan Shu  (corresponding author), \href{mailto:yizhans@princeton.edu}{yizhans@princeton.edu}, 
Chenyu Yu, \href{mailto:chenyu@princeton.edu}{chenyu@princeton.edu}, 
John M. Mulvey, \href{mailto:mulvey@princeton.edu}{mulvey@princeton.edu} 
}
}

\date{August 24, 2024}

\begin{document}

\maketitle


\begin{abstract}

This article investigates a regime-switching investment strategy aimed at mitigating downside risk by reducing market exposure during anticipated unfavorable market regimes. 
We highlight the \emph{statistical jump model} (JM) for market regime identification, a recently developed robust model that distinguishes itself from traditional Markov-switching models by enhancing regime persistence through a jump penalty applied at each state transition. 
Our JM utilizes a feature set comprising risk and return measures derived solely from the return series, with the optimal jump penalty selected through a time-series cross-validation method that directly optimizes strategy performance.           %
Our empirical analysis evaluates the realistic out-of-sample performance of various strategies on major equity indices from the US, Germany, and Japan from 1990 to 2023, in the presence of transaction costs and trading delays.
The results demonstrate the consistent outperformance of the JM-guided strategy in reducing risk metrics such as volatility and maximum drawdown, and enhancing risk-adjusted returns like the Sharpe ratio, when compared to both hidden Markov model-guided strategy and the buy-and-hold strategy. 
These findings underline the enhanced persistence, practicality, and versatility of strategies utilizing JMs for regime-switching signals.

\end{abstract}


\newpage
\section{Introduction}

Historically, traded markets have exhibited cyclical and time-varying behavior, driven by a complex interplay of economic, behavioral, and political factors. 
A \emph{financial regime} is defined by extended, consecutive periods displaying homogeneous market behavior, manifesting as bull or bear markets characterized by positive or negative returns \citep{Pagan2003}, calm or turbulent periods marked by low or high volatility \citep{Schwert1989}, and risk-on or risk-off episodes reflecting market sentiment \citep{smales2016}, among other patterns. 
A regime switch denotes a sudden yet persistent shift in these behaviors.
Evidence of regime-switching patterns has been established across various asset classes, including equities \citep{hardy2001}, fixed income \citep{gray1996}, and currencies \citep{REUS2016}. 
A key advantage of regime-switching models is their interpretability; the identified regimes often align with real-world events like different phases of the macroeconomic business cycle \citep{hamilton1989}. 
These models not only correspond with shifting fundamentals -- typically understood only ex-post -- but also enable ex-ante real-time online inference \citep{ang2012}.   %

A principal application of financial regimes lies in mitigating downside risk -- the potential for losses under adverse market conditions\footnote{
Here, ``downside risk'' refers to the risk of downside losses, relative to upside gains, under unfavorable conditions. This term is distinct from the stock-level downside risk factor that measures a security’s sensitivity to negative market movements, as proposed in \citet{Ang2006}.
}.
Among the various identified regimes, there often exists one marked by increased variances and covariances alongside lower, frequently negative returns. 
Commonly termed ``bad times'', ``bear markets'' or ``volatile/high volatility regime”, this characterization is prevalent among both academics and financial media. 
As foundational works in regime-based asset allocation \citep{ang2002asset, ang2004} suggest, when a bear market is anticipated, the optimal strategy typically involves shifting to safer assets, such as the risk-free asset. 
This is because, ``in this regime, interest rates tend to be on average higher and equity returns more negatively correlated with the short rate''.
Consequently, maintaining significant exposure to risky assets during bear markets due to the ignorance of regime-switching dynamics can lead to substantial economic costs, including wealth losses and drawdowns resulting from inadequate downside risk management.

In this article, we explore a straightforward regime-switching investment strategy inspired by \citet{bulla2011strategy} to mitigate downside risk,  with a specific focus on comparing the out-of-sample performance of various regime identification models under realistic conditions. 
This strategy, referred to as the ``\mbox{0/1} strategy'', is applied to an individual asset and utilizes its asset-specific regime identification by allocating 100\% to the asset if a bull market is anticipated, or conversely, shifting 100\% of the investment to the risk-free asset in anticipation of a bear market  -- hence the strategy's name.
The simplicity of this strategy makes it an effective tool for evaluating the financial implications of the accuracy of different regime identification models.

We consider regime identification models\footnote{
The term \emph{regime identification models}, also commonly known as \emph{regime-switching models} in the literature, is used here to emphasize their ability to identify unobserved regimes.
}
from both parametric and non-parametric families, with hidden Markov models (HMMs) and statistical jump models (JMs) as prominent representatives of each family.
The ability of HMMs to reduce risk and enhance risk-adjusted return when applied to the 0/1 strategy has been substantiated in \citet{bulla2011strategy}, and thus they serve as the benchmark model in our study.
Despite the long research history of Markov-switching models, recent studies have highlighted the sensitivity of HMMs to model mis-estimation and mis-specification \citep{nystrup2020jump, nystrup2020online}. 
Mis-estimation issues may arise from limited sample sizes, unbalanced data, or high state persistence, while model mis-specification often results from the non-normal and time-varying distributions, characteristics all typical and inherent to financial return series.

In response to the challenges outlined above, various non-parametric regime identification models have recently emerged, with our study focusing particularly on statistical jump models (JMs)\footnote{
The JMs discussed in this article are not related to \emph{jump-diffusion models}, commonly used in stochastic finance. 
Throughout this article, we will use the terms statistical jump model, jump model, and JM interchangeably.
}.
JMs are based on the principle of clustering temporal features while simultaneously imposing penalties for each jump in the hidden state sequence.
This penalization is regulated by a hyperparameter known as the \emph{jump penalty}, which reflects our prior belief about the desired persistence of the hidden state sequence.
In our approach, we design our JM employing a simple feature set comprising risk and return measures derived solely from the asset return series, ensuring a fair comparison with HMMs, which utilize daily return series.
Building on previous research that has demonstrated the enhanced persistence of inferred state sequences and improved identification accuracy in JMs, our primary contribution is to illustrate the financial benefits of these enhancements through a direct application in a regime-switching investment strategy.

Another contribution of our work lies in showcasing the optimal jump penalty selection in a realistic live-sample setting via a time-series cross-validation approach.
As previously noted, the jump penalty moderates the frequency of state transitions, and more intrinsically relates to the tradeoff between accuracy and latency, a common challenge in regime-switching strategies as discussed in \citet{Nystrup2018JPM}.
Unlike earlier methods that select jump penalties based primarily on statistical criteria, our approach chooses the jump penalty that achieves the highest Sharpe ratio of the JM-informed 0/1 strategy over the validation period.
This method aims to directly maximize the practical benefits for financial applications that rely on the regimes inferred by JMs.

In our empirical analysis, we evaluate the out-of-sample performance of the 0/1 strategy using regimes inferred by JMs and HMMs, comparing it to the buy-and-hold strategy, when applied individually to daily equity indices from the US, Germany, and Japan.
We incorporate realistic conditions, including transaction costs and a one-day trading delay, across a testing period from 1990 to 2023.
Our results reveal that the 0/1 strategy, particularly when informed by JMs, significantly reduces risk metrics like volatility and maximum drawdown, highlighting its effectiveness in mitigating downside risk.
The JM-informed strategy improves annualized returns by approximately 1\% to 4\% across different regions, leading to improved risk-adjusted return metrics.
Additionally, the inherent persistence in JMs provides enhanced robustness against trading delays. 
Although this article focuses on equity indices, our methodology is versatile enough to encompass various asset classes and can be adapted for the return series of specific trading strategies, such as trend following or long-short factor portfolios\footnote{
In this context, the regime identification model aims to discern periods of strong or weak performance for the trading strategy, with the 0/1 strategy acting to timely activate or deactivate the trading strategy based on prevailing market conditions.
An example is provided by \citet{Bosancic2024}, who are inspired by our article to apply the 0/1 strategy to a collection of long-only factor portfolios. 
(Our original manuscript was publicly released prior to the start of their work.)
}.

The outline of the article is as follows.
We begin with an overview of the data used in our empirical analysis in Section \ref{sec:data}.
Section \ref{sec:method} delves into our methodology and discusses related literature, providing a detailed description of the regime-switching investment strategy and the regime identification models. 
This section also includes some results from model fitting.
In Section \ref{sec:results}, we present the empirical findings, comparing the out-of-sample performance of the strategies and illustrating the inferred regimes. 
The article concludes in Section \ref{sec:conclusion}.

\section{Data}   \label{sec:data}

The data analyzed in this article comprises the daily total return series of three major equity indices: \mbox{S\&P 500}, DAX, and \mbox{Nikkei 225}, representing the US, Germany, and Japan, respectively.
These data are sourced from the Bloomberg Terminal\footnote{
All indices are denominated in their local currencies. We avoid the consideration of exchange rate risk.
}.
For the risk-free rates, we use the \mbox{3-month} Treasury Bill Yield from each corresponding country, sourced from the Global Financial Data (GFD) database.
All data spans from the start of 1970 to the end of 2023. 
It is important to note that our methodology is tested independently on each index, rather than constructing an international equity portfolio that combines all three indices.

\begin{figure}[htb]
    \centering
    \includegraphics[width=\textwidth]{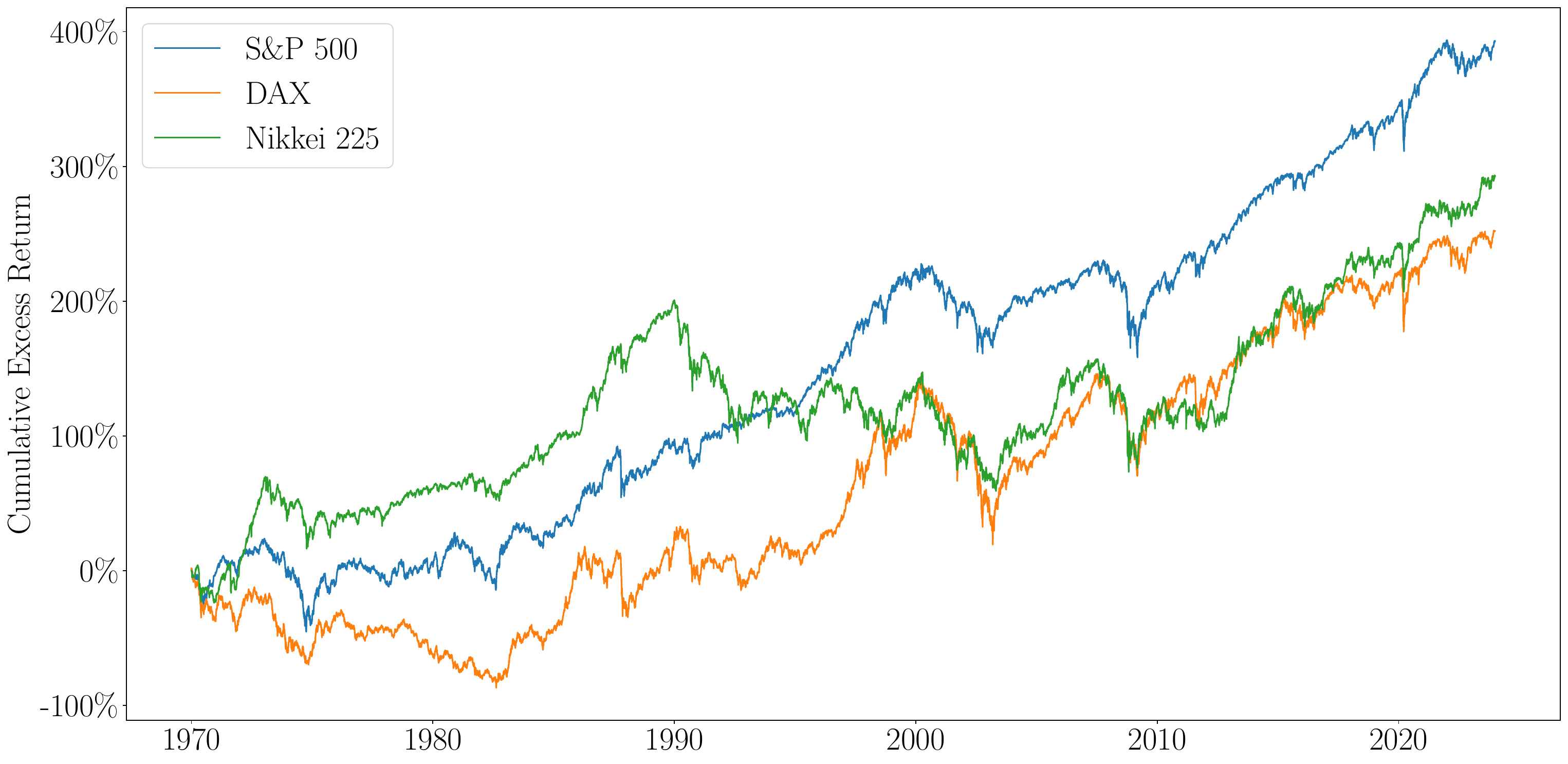}
    \caption{\small Cumulative excess returns of the S\&P 500, DAX, and Nikkei 225 indices from 1970 to 2023.}
    \label{fig:index cumret}
\end{figure}

Figure \ref{fig:index cumret} presents the cumulative excess returns for the three indices from 1970 to 2023. 
This visualization reveals distinct patterns among the indices.
The S\&P 500 shows the highest total excess return, reflecting the consistent economic growth and stability of the US market over the analyzed period. 
The DAX, on the other hand, exhibits more pronounced volatility with deeper drawdowns during turbulent periods. 
The Nikkei 225 illustrates the economic downturn in Japan, especially the prolonged stagnation following the asset bubble burst in the early 1990s. 
Throughout the period, all indices display persistent bear and bull markets, with significant losses during major events such as the dot-com bubble, the 2008 financial crisis, and the 2020 COVID-19 crash.
These patterns highlight the value of regime-switching models to generate signals that can help mitigate downside risk by shifting to safer assets during extended bear markets.
Despite a few extreme returns during these events, we do not process outlier values to minimize manual intervention.

The diverse behaviors of these indices are further supported by their relatively low correlations, as shown in Table \ref{tab:data_corr}.
The S\&P 500 and DAX have the highest correlation at 0.44, while the \mbox{S\&P 500} and Nikkei 225 have the lowest correlation at 0.12. 
Despite generally low long-run correlations, it is important to note that international correlations tend to increase during bear markets defined by large negative returns, as formally confirmed by statistical tests in \citet{Longin2001}.
The diverse behaviors, along with the low correlations, highlights the international relevance and broad applicability of our methodology.

\begin{table}[tb]
    \centering

    \begin{tabular}{cccc}
\toprule
 & S\&P 500 & DAX & Nikkei 225 \\
\midrule
S\&P 500 & $\lp{17.2\%}^2$ & \cellcolor{gray!30}0.44 & \cellcolor{gray!30}0.12 \\
DAX & 1.5\%  & $\lp{20.1\%}^2$ & \cellcolor{gray!30}0.26 \\
Nikkei 225 & 0.4\%  &  1.1\% & $\lp{20.5\%}^2$ \\
\bottomrule
\end{tabular}

    \caption{\small Annualized variances, covariances, and correlations of daily excess returns for the S\&P 500, DAX, and Nikkei 225 indices from 1970 to 2023.
    Variances are shown along the diagonal, covariances in the lower diagonal, and correlations in the upper diagonal (shaded). Covariances and correlations are calculated using data from common trading days.}
    \label{tab:data_corr}
\end{table}

\section{Methodology}   \label{sec:method}

In this section, we first discuss the 0/1 strategy, followed by an overview of the regime identification models that generate switching signals for this strategy. 
We then provide detailed discussions on two specific models: hidden Markov models (HMMs) and statistical jump models (JMs), including observations from model fitting results.

\subsection{Regime-Switching Investment Strategy}

We focus on the 0/1 strategy, which switches between a 100\% investment in either a risky asset or a risk-free asset, based on whether the inferred prevailing regime for the risky asset is favorable, such as a bull market or low volatility regime.
The economic rationale is straightforward: we shift to safer assets when a persistent unfavorable regime is anticipated. 
This binary asset universe -- comprising one risky and one risk-free asset -- has long been studied in regime-based asset allocation. 
For instance, \citet{Guidolin2004} explore the optimal portfolio choice within these two assets under a regime-switching return process, while \citet{Nystrup2018QF} address multi-period portfolio optimization in this universe under changing market conditions. 
By isolating the potential for diversification, the performance of the 0/1 strategy offers a clear evaluation of the added value from asset-specific regime identification, aligning perfectly with our goal to compare the out-of-sample performance of various regime identification models. 
On the other hand, numerous studies, including those by \citet{vliet2011, sheikh2012}, have investigated \emph{dynamic asset allocation} that spans a broader range of asset classes to leverage diversification benefits. 
Specifically, \citet{shu2024AA} demonstrate a systematic approach using various portfolio construction methods to integrate asset-specific regime identification, as generated by JMs in this article, into a multi-asset portfolio.

Alternative regime-based weight allocation methods exist within this asset universe, yet they often perform similarly to the simpler 0/1 weighting scheme. 
For instance, \citet{nystrup2015JPM} test a long-short strategy on an equity index based on identified regimes. 
While this approach enhances returns, the shorting positions fail to effectively mitigate risk, resulting in high volatility and drawdowns. 
Furthermore, \citet{nystrup2016changePoint} show that gradually adjusting the weight on the equity index as a linear function of the forecasted probability delivers similar out-of-sample performance compared to switching allocations between 100\% and 0\%. 
It's worth noting that such a drastic change in allocation may be too extreme for practical trading applications. 
Although not explored in this work, this switching signal could be incorporated into a static benchmark portfolio in various ways.
To enhance practical applicability, the persistence of regime forecasts emerges as a critical factor in preventing excessive rebalancing while maintaining identification accuracy, an aspect where jump models can offer advantages.
The importance of the persistence of inferred regimes is further discussed at the end of Section \ref{subsec:overview RI models}.

A practical issue that may arise is the delay in trading the signal. 
Ideally, a no-delay scenario would allow portfolio managers to apply the switching signal generated at the end of day $t$ directly on the next trading day $t+1$. 
However, this immediacy might be challenging for managers overseeing large investments. 
Following previous studies \citep{nystrup2015JPM, Nystrup2018JPM}, we assume a one-day delay, meaning that if the prevailing regime inferred at the end of day $t$ differs from the regime on which the current allocation is based, then the switch in allocation can only be executed at the end of next day $t+1$, or, in other words, the signal applied starting from day $t+2$. 
To assess the impact of this delay, we conduct a robustness test in Section \ref{subsec:sensitivity}, where we present results for delays of up to two weeks to evaluate the strategy's robustness.

\subsection{Overview of Regime Identification Models}   \label{subsec:overview RI models}

\citet{harding2003} sparked a debate over the advantages and disadvantages of parametric versus non-parametric models in applying regime identification models to date business cycles \citep{hamilton2003, harding2003rejoinder}. 
Parametric models assume specific probability distributions for observations and use estimated parameters to infer regime assignments. 
In contrast, non-parametric models adopt a likelihood-free, data-driven approach that focuses on directly identifying the unobserved regime sequence, particularly the turning points. 
We contribute to this discussion by evaluating the out-of-sample performance of two representative models from each family.

Markov-switching models form a substantial portion of the parametric family, integrating regime-switching dynamics into various models, including Gaussian distributions \citep{turner1989}, autoregressive models \citep{hamilton1989}, and ARCH models \citep{hamilton1994}, to characterize both macroeconomic variables and financial returns.
At their core, these models utilize an unobserved finite-state Markov chain that governs the parameters of the distribution from which observations are generated.
This chain is often characterized by high diagonal values in the transition probability matrix, close to 1.0, reflecting significant state persistence.
For our purposes, hidden Markov models (HMMs) have been demonstrated by \citet{ryden1998} to adequately capture many temporal and distributional stylized facts in equity index returns, and by numerous studies such as \citet{bulla2011strategy, bae14, nystrup2015JPM} to enhance investment performance in regime-switching strategies.    %
Therefore, we adopt the original Gaussian HMMs as our benchmark parametric model.
Further developments in HMMs include, for example, using heavy-tailed conditional distributions \citep{bulla2011tHMM} and sojourn time distributions different from the geometric distribution \citep{bulla06}.

Parallel advancements have been made in non-parametric regime identification models, aimed at addressing the limitations of the parametric family, including restriction to predetermined distributional assumptions, challenges in computing maximum likelihood estimation (MLE) due to the presence of latent variables, and a lack of transparency in the identification rules. 
Initially, the non-parametric business cycle dating algorithm in \citet{harding2003} comprised a set of simple rules coupled with persistence-enhancing censoring procedures, designed specifically for univariate time series.
Over time, more sophisticated techniques have emerged, many of which incorporate insights from  machine learning algorithms such as trend filtering \citep{mulvey2016, kim2023}, spectral clustering \citep{zheng2021}, $k$-means clustering \citep{Akioyamen2021, Greevy2024}, and recurrent neural networks \citep{fantulin2024}. 
Among these alternatives, we focus on statistical jump models that enhance $k$-means clustering by imposing a jump penalty with each regime switch, offering improved regime persistence and a potential for application in the 0/1 strategy that involves significant portfolio rebalancing.

Before examining each of the two models in detail, it is useful to highlight some of their commonalities. 
First, both models infer regimes directly from market data -- specifically, daily equity index returns -- rather than macroeconomic indicators, as utilized in many studies \citep{kim2023, Elkamhi2023}. 
While those studies focus on \emph{economic regimes} that influence all considered assets or regions through broad economic conditions, our research aims to identify \emph{financial regimes} tailored to specific markets based on asset returns.
Second, both models employ a two-state implementation ($K=2$), which, despite statistical criteria often favoring more states \citep{Guidolin2011} (typically no more than four), has proven sufficient for capturing the dynamics of a single return series\footnote{
As noted by \citet{Guidolin2006}, ``while simple two- or three-state models capture the univariate dynamics in [a single asset return series], a more complicated four-state model $\ldots$ is required to capture the joint distribution of [multiple assets]''.
}, and provides better model interpretability and stability. 
Nevertheless, there are differences: HMMs characterize regimes solely by conditional volatilities, whereas JMs consider a broader set of features including both return and risk measures. 
In our notation, $s_t=0$  denotes the favorable regime, and $s_t=1$ denotes the unfavorable one.
Third, both models generate daily switching signals to alleviate the impact of incorrect regime forecasts from a whole month to shorter spans.

Finally, it is critical to emphasize the unsupervised nature of regime identification models, which stems from the unobserved nature of the regimes themselves\footnote{
While observed regimes, such as the business cycles dated by the NBER, exist, their direct application for investment purposes is limited due to substantial data availability lags, typically ranging from 4 to 21 months as reported in \citet{Giusto2017}, and infrequent updates.
}. 
Specifically, the model is trained to discern the unobserved state $s_t$ for day $t$ based on features $\bmx_t$ available at the end of that day.
This contrasts with a supervised forecasting exercise where, for example, all available information at the end of day $t$ is used to predict  asset returns for the next day $t+1$.
In this context, regime identification models are interpretative rather than predictive, focusing on understanding daily market dynamics\footnote{
Thus, in this article, we avoid terms like ``regime prediction/forecasting''.
The task of using all available information at the end of day $t$ to predict the market regime $s_{t+1}$ tomorrow is explored in studies such as \citet{james19, Uysal2021, shu2024AA}, which employ classification algorithms including support vector machines and (gradient-boosted) decision trees.
The prediction label $s_{t+1}$ may correspond to observed regimes or unobserved ones inferred from regime identification models, as described in this article.
}.
The switching signal generated by these models at the end of day $t$, indicating the prevailing regime, usually consists of the identified regime $\hat s_t$ (potentially subject to smoothing) for day $t$ as the last day of the latest training window containing features $\bmx_t$\,\footnote{
Although Markov-switching models theoretically have predictive capabilities due to the Markov chain assumption, the high diagonal values in the transition probability matrix, especially in daily models, suggest that the predicted state for day $t+1$ mostly remains the same as the identified state for day $t$.
}.
Consequently, the profitability of regime-based strategies heavily relies on the assumption of regime persistence, hoping that the regime identified for day $t$ will persist into the following days and beyond.
As \citet{nystrup2015JPM} summarize, the purpose of regime-based strategies is ``not to predict regime shifts or future market movements, but to identify when a regime shift has occurred, and then benefit from the persistence of equilibrium risk-return relations''.
This, in turn, places a high demand on the persistence of regimes inferred from the models, as a regime signal that flips very frequently is unlikely to be useful for investment purposes given the constraint that portfolio managers can only change allocation from day $t+2$ or even later, based on the regime inferred for day $t$.
This delay between identification and trading implementation can easily offset the potential profits from a rapidly changing signal.
Differences in persistence between HMMs and JMs are detailed in \mbox{Table \ref{tab:num shifts}} and Section \ref{sec:results}.

\subsection{Hidden Markov Models}

We conduct a rolling fit of a two-state Gaussian HMM\footnote{
For brevity, we omit a detailed description of HMMs. Interested readers may refer the monograph by \citet{Zucchini2016}.
} 
on daily log total returns, using a training window of 3000 days (approximately 12 years) that moves forward daily, following the practice in \citet{bulla2011strategy}.  
Our implementation uses the Python package \texttt{hmmlearn}.
To address the non-convex nature of the likelihood function, we execute the algorithm ten times from different initial values of conditional returns derived from the \mbox{$k$-mean$++$} algorithm, and retain the fitting that yields the highest log-likelihood.
The two regimes are distinguished by their estimated conditional volatilities, categorized as either low or high volatility regimes.

\begin{figure}[htbp]
    \centering
    \includegraphics[width=\textwidth]{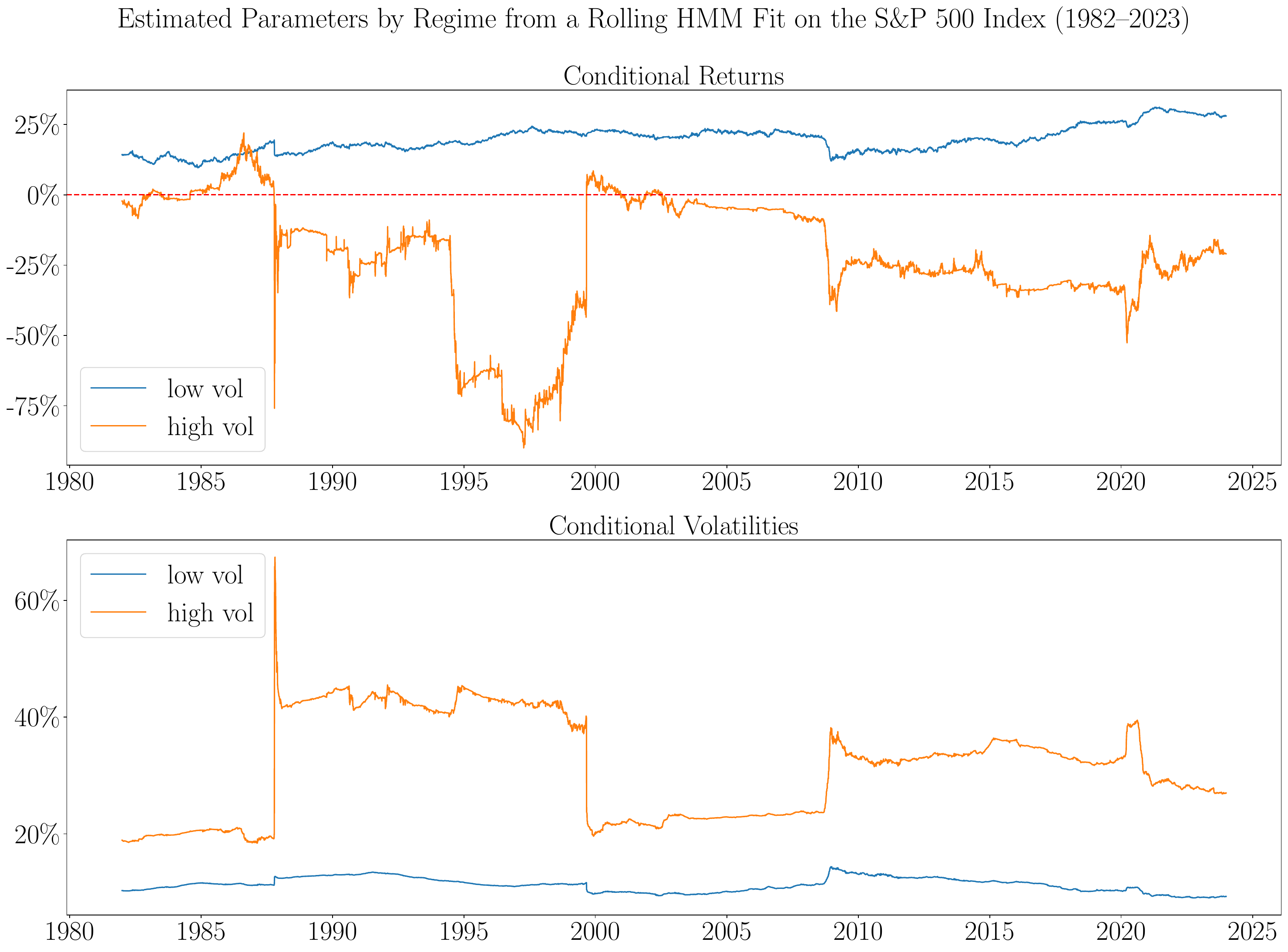}
    \caption{\small Estimated conditional returns and volatilites by regime from a rolling HMM fit on the S\&P 500 index from 1982 to 2023. 
The HMM is fitted using a 3000-day training window that moves forward daily.
Each date on the $x$-axis represents the end of a training window.
All values on the $y$-axis are annualized.
The yellow curve represents the estimated parameters for the high volatility regime, with the blue curve representing the low volatility regime.
}
    \label{fig:hmm_params}
\end{figure}

Figure \ref{fig:hmm_params} illustrates the evolution over time of the estimated conditional returns and volatilites (annualized) for each regime from the rolling HMM fit on the S\&P 500 index\footnote{
For brevity, in this section, we present the model fitting results only for the S\&P 500 index. 
The main conclusions are consistent across the other two indices.  
Comprehensive out-of-sample performance of the 0/1 strategy for all three indices is presented in Section \ref{sec:results}.
} 
from 1982 to 2023.
Each date on the $x$-axis marks the end of a training window.
The estimated parameters substantiate many previously reported observations.
The regimes are consistently differentiated by volatilites, with the estimated volatilities under the two regimes differing by at least a factor of two.
Furthermore, the high volatility regime predominantly experiences negative returns, reflecting the lower risk-adjusted returns during turbulent periods.

On the training window ending at day $t$, we use the Viterbi algorithm to decode the hidden state sequence and extract the last identified state $\hat s_t$ as the online inferred prevailing regime. 
As noted in \citet{bulla2011strategy}, identification error rates peak at the start and end of the training window, posing a significant challenge for our online inference application.
A consequence is that the persistence of the online inferred state sequence $\lp{\hat s_t}$ is compromised. 
Following the solution in \citet{bulla2011strategy}, we apply a median filter, equivalent to computing a rolling mean on the sequence $\lp{\hat s_t}$  over a trailing window of length $k$, and only signal a high volatility regime when the rolling mean exceeds 0.5 (recall that low/high volatility regimes are denoted by $s_t=0/1$, respectively).
Originally, \( k \) was set at 6; in our approach, it is selected from a range of candidate values automatically via a cross-validation framework.
This same method, detailed in Section \ref{subsec:hyper tuning}, is also used to select the optimal jump penalty for JMs.

\subsection{Statistical Jump Models}

\citet{Bemporad2018} introduce statistical jump models (JMs) as a general unsupervised learning algorithm that fits multiple model parameters to time series data while incorporating temporal information. 
In our application, given an observation sequence of $D$ standardized features $\bmx_0,\ldots,\bmx_{T-1}\in \R^D$ derived from a return series $r_0,\ldots, r_{T-1}$ (features detailed in Section \ref{subsec:feature}), we estimate a two-state JM ($K=2$) by solving the following optimization problem:
\begin{equation}\label{eq:jumpObj}
\min_{\Theta, \bmS}\quad\sum_{t = 0}^{T-1} l(\bmx_t, \bmtheta_{s_t})+\lambda\sum_{t = 1}^{T-1} \indicator{s_{t-1} \neq s_{t}}	\,. 
\end{equation} 
Here, the optimization variables are the $K$ model parameters $\Theta:= \lB{\bmtheta_k \in\R^D :k= 0, \ldots, K-1}$, and the unobserved state sequence $\bmS:=\{s_0,\ldots, s_{T-1} \}$. 
Each model parameter $\bmtheta_k$ serves as a centroid for the \mbox{$k$-th} state, and each state variable $s_t$ takes one of the $K$ discrete values in $\{0, 1, \ldots, K-1\}$, representing the assigned state for day $t$ and linking features $\bmx_t$  with the corresponding model parameter $\bmtheta_{s_t}$, whose dissimilarity is quantified by a loss function $l(\cdot, \cdot)$, here chosen as the scaled squared $\ell_2$-distance, defined by  $l(\bmx, \bmtheta) := \frac12\lV[2]{\bmx- \bmtheta}^2$.
After estimating the JM, we derive the transition probability matrix from the optimal hidden state sequence and calculate metrics such as return and volatility for each state.
We distinguish between bull and bear market states based on the cumulative excess return for each state over the training period.
The state with the higher cumulative return is identified as the bull regime ($s_t=0$), while the state typically exhibiting negative cumulative return is considered bearish ($s_t=1$).

The objective function \eqref{eq:jumpObj} balances fitting the data with multiple model parameters and incorporating our prior beliefs about the persistence of the state sequence, where $\lambda\ge0$ is a critical hyperparameter known as the \emph{jump penalty}.
This penalty moderates the fixed-cost regularization term, incurred whenever there is a jump between two consecutive state variables.
With a zero jump penalty, the model reduces to a $k$-means clustering algorithm, ignoring temporal information. 
As $\lambda$ increases, state transitions become less frequent, eventually grouping all data points into a single cluster if $\lambda$ is high enough.
The optimal selection of the jump penalty is discussed in \ref{subsec:hyper tuning}.

To minimize the objective function \eqref{eq:jumpObj}, a coordinate descent algorithm outlined in \citet{nystrup2020jump} is used, alternating between optimizing the model parameters $\Theta$ and the state sequence $\bmS$  while keeping the other variable fixed from the previous iteration\footnote{
An implementation of a collection of jump models is available on the first author’s GitHub page (\url{https://github.com/Yizhan-Oliver-Shu/jump-models}).
}. 
Optimization over $\Theta$ is computed within each state, while optimization over the discrete variable sequence $\bmS$ can be approached via a dynamic programming (DP) algorithm in linear time.
This DP algorithm is crucial for performing online inference in a real-time live-sample setting, as described in \ref{subsec:online inference}.
We run the algorithm ten times and retain the fitting with the lowest objective value, using the same approach as described for HMM fitting.

\citet{Bemporad2018} demonstrate that, under certain assumptions, JMs nest HMMs in a probabilistic setting. 
Subsequently, \citet{nystrup2020jump} introduce JMs to financial applications, illustrating that JMs can outperform HMMs in identification accuracy and state sequence persistence.   
\citet{Aydinhan2024} expand JMs into the \emph{continuous statistical jump model} (CJM) by generalizing the discrete hidden state variable into a probability vector across all states, offering a probabilistic interpretation where the hidden state vector represents the probability of each period belonging to each regime. 
While these probability values hold potential use, the discrete nature of the 0/1 weighting scheme we employ here shows no significant difference in strategy performance between JM and CJM, based on our experiences.

\subsubsection{Features}   \label{subsec:feature}

For JMs, we employ a feature set outlined in Table \ref{tab:JM features}, consisting of three return and risk measures derived from an excess return series. 
These features are applied across all three equity indices to ensure robustness without further customization. 
Specifically, we use an exponentially weighted moving (EWM) downside deviation (DD) with a halflife of 10 trading days, and EWM Sortino ratios with halflives of 20 and 60  days.
These halflife values are adaptable for applications to other asset classes.

\begin{table}[htbp]
    \centering
    \begin{tabular}{c c c c}
    \toprule
         \textbf{No.} & \textbf{Category} & \textbf{Feature} & \textbf{Halflives (days)} \\
         \midrule
          1 & Risk & Downside Deviation & 10 \\  
         \midrule
         2 & Return & Sortino Ratio & 20 \\ 
         \midrule
         3 & Return & Sortino Ratio & 60 \\
          \bottomrule 
          \multicolumn{4}{l}{\rule{0pt}{2.75ex} Note: Each feature is exponentially smoothed.}
    \end{tabular}
    \caption{\small List of features used in jump models, exponentially smoothed over specified halflives, derived from an excess return series.}
    \label{tab:JM features}
\end{table}

Downside deviation, calculated as $\E{R^2\bfone_{\{R<0\}}}$ where $R$ denotes the excess return and the expectation is based on exponentially decaying weights over historical periods, is our preferred risk measure. 
This preference stems from the observation in early works, such as the model by \citet{Roy1952} and the ``semi-variance'' proposed by \citet{Markowitz1959}, that investors are more concerned with downside losses than the uncertainty of upside gains. 
Given the relatively smooth evolution of risk measures over time, we choose a shorter halflife of 10 days.
For the return measure, we opt for the EWM Sortino ratio, calculated as the ratio of EWM average returns to EWM DD, each computed over a specified halflife, to assess risk-adjusted returns. 
Considering the high level of noise typically associated with return measures, we incorporate EWM Sortino ratios over two longer halflives -- 20 and 60 days -- to mitigate the impact of rapidly changing features which could result in too frequent state changes in the inferred state sequence, even under significant jump penalties.
To maintain a balanced emphasis on both risk and return aspects -- unlike HMMs which focus solely on volatility -- we include two return measures and one risk measure, taking into account the substantial noise in return measures.

To ensure a fair comparison with HMMs, which utilizes the daily return series itself, this article develops only a parsimonious set of features for JMs.
Although JMs can perform similarly to HMMs when given the return series itself as a one-dimensional input, the true advantage of JMs lies in their capacity to integrate a more extensive feature set, as compared in \citet{nystrup2021sparse} for high-dimensional HMMs and JMs. 
When considering an expanded feature set, the \emph{sparse statistical jump model} proposed by \citet{nystrup2021sparse} could be of potential utility. 
This model performs feature selection based on the in-sample clustering effect of each feature, with applications in identifying cryptocurrency market regimes  \citep{cortese2023crypto}. 
Alternatively, \citet{Bosancic2024} suggest a heuristic for feature subset selection. 
Additional exogenous features such as macroeconomic indicators \citep{kim2023} and cross-asset variables like stock-bond correlation \citep{Yang2009} could be included into the feature set of JMs.

\begin{figure}[htbp]
    \centering
    \includegraphics[width=\textwidth]{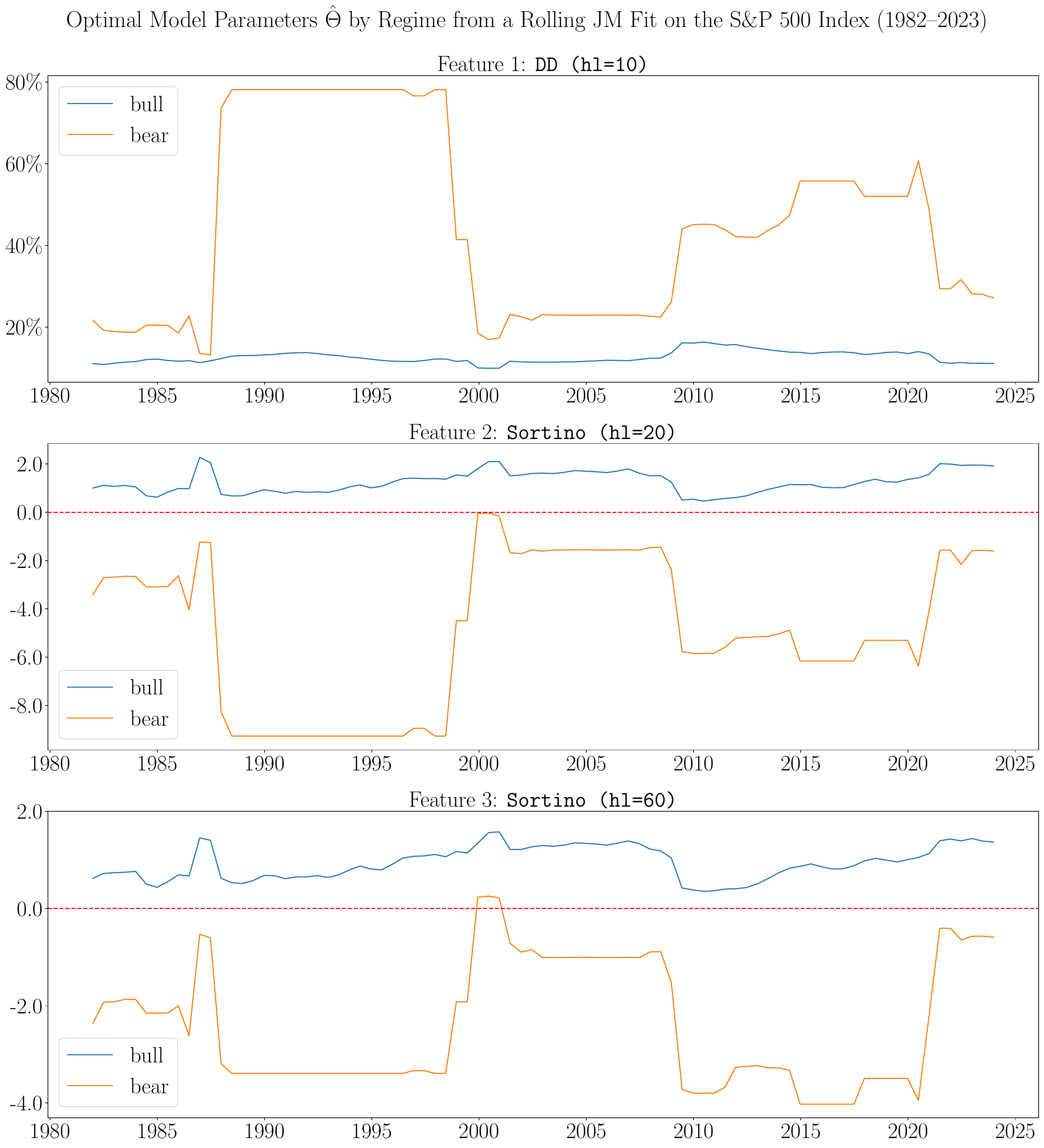}
    \caption{\small Optimal model parameters for all three features (referenced in Table \ref{tab:JM features}) by regime from a rolling JM fit on the S\&P 500 index from 1982 to 2023.
The JM uses a fixed jump penalty of 50.0 and is fitted using a 3000-day training window that moves forward every six months.
Each date on the $x$-axis represents the end of a training window.
All values on the $y$-axis are annualized; the downside deviation feature is scaled by $\sqrt 2$, and the Sortino ratio features are divided by $\sqrt 2$, to ensure consistency with volatility and Sharpe ratio calculations.
The yellow curve represents the optimal model parameters for the bear regime, with the blue curve representing the bull regime.
}
    \label{fig:jm_params}
\end{figure}

In our JM implementation, the optimal model parameters $\hat\Theta$ are updated every six months by solving the full optimization problem in equation \eqref{eq:jumpObj} over a 3000-day training window.
We fit the JM at this reduced frequency to ensure the stability and consistency of the optimal $\hat\Theta$.
Figure \ref{fig:jm_params} illustrates the evolution over time of the optimal $\hat\Theta$ (annualized) for each regime from the rolling JM fit on the S\&P 500 index from 1982 to 2023, serving as a counterpart to the parameter plot for HMMs in Figure \ref{fig:hmm_params}.
The jump penalty is fixed at a typical value $\lambda=50.0$. 
Under the bull regime, the optimal features are characterized by low risk and high return measures, while the bear regime features elevated risk and negative return measures.

\subsubsection{Online Inference}  \label{subsec:online inference}

An online algorithm processes its input data sequentially, observation by observation, as soon as the latest data becomes available. 
In our application, under a fixed jump penalty, with the most recently updated optimal model parameters $\hat \Theta$, we need to infer the state to which \mbox{day $t$} belongs based on the features $\bmx_t$ available at the end of that day, without access to future data $\bmx_{t+1}$ of tomorrow and beyond. 
A basic $k$-means style online inference would involve directly assigning features $\bmx_t$ to the nearest centroid, \mbox{i.e.}, $\hat{s}_t = \argmin_kl(\bmx_t, \hat{\bmtheta}_{k})$, which ignores temporal information.
To enhance the persistence of the online inferred regime sequence, we incorporate a lookback window $\bmx_{t-l+1}, \ldots, \bmx_t$ with $l=3000$, the same as the training window length, and run the dynamic programming (DP) algorithm to minimize the objective function \eqref{eq:jumpObj} over the state sequence $s_{t-l+1}, \ldots, s_t$, while keeping the model parameters fixed at the most recent optimal values $\hat{\Theta}$.
We then extract the last state $\hat{s}_t$ from the optimal state sequence as the online inferred prevailing regime.
\citet{nystrup2020online} describe a computationally efficient method for performing the above calculations for a range of consecutive days.

The above procedure for generating online regime inference is applicable to any fixed jump penalty.
Here, we illustrate the differences in persistence between HMMs and JMs, and how it is influenced by their respective smoothing hyperparameter: the window length $k$, over which the rolling mean of $(\hat s_t)$ is calculated for HMMs, and the jump penalty $\lambda$ for JMs.
In Table \ref{tab:num shifts}, we present the average number of regime shifts per year in the online inferred state sequence from 1982 to 2023, generated by HMMs and JMs under fixed smoothing hyperparameters.
For HMMs, increasing $k$ can reduce the number of regime shifts, but not very efficiently. 
Even with $k=20$, which corresponds to a signal latency of at least 10 days, the average number of shifts remains at about 2 per year.
In contrast, adjusting $\lambda$ effectively smooths the inferred sequence, reducing the number of shifts from 9.7 times under no jump penalty (equivalent to $k$-means clustering) to 2.7 times with a very small penalty $\lambda=5.0$. 
At a typical value between 50.0 to 100.0, the number of shifts per year is reduced to fewer than one, aligning with the frequency in observed regimes such as the business cycle.

\begin{table}[htbp]
    \centering
\begin{tabular}{l*{11}{c}}
\toprule
&\multicolumn{5}{c}{HMM} & \multicolumn{6}{c}{JM}  \\
    \cmidrule(lr){2-6} \cmidrule(lr){7-12}
    
 & $k=0$ & 2 & 4 & 8 & 20 & $\lambda=0.0$ & 5.0 & 15.0 & 35.0 & 70.0 & 150.0 \\
\midrule
\# of shifts & 8.5 & 6.6 & 4.9 & 3.2 & 2.0 & 9.7 & 2.7 & 1.7 & 0.8 & 0.5 & 0.4 \\
\bottomrule
\end{tabular}
    \caption{\small Average number of shifts per year in the online inferred regime sequence from 1982 to 2023, generated by HMMs, smoothed by taking rolling means over different lookback window lengths $k$, and by JMs, using different jump penalty values $\lambda$.
    The hyperparameter $k$ and $\lambda$ are fixed throughout the period.
    }
    \label{tab:num shifts}
\end{table}

\begin{figure}[tb]
    \centering
    \includegraphics[width=\textwidth]{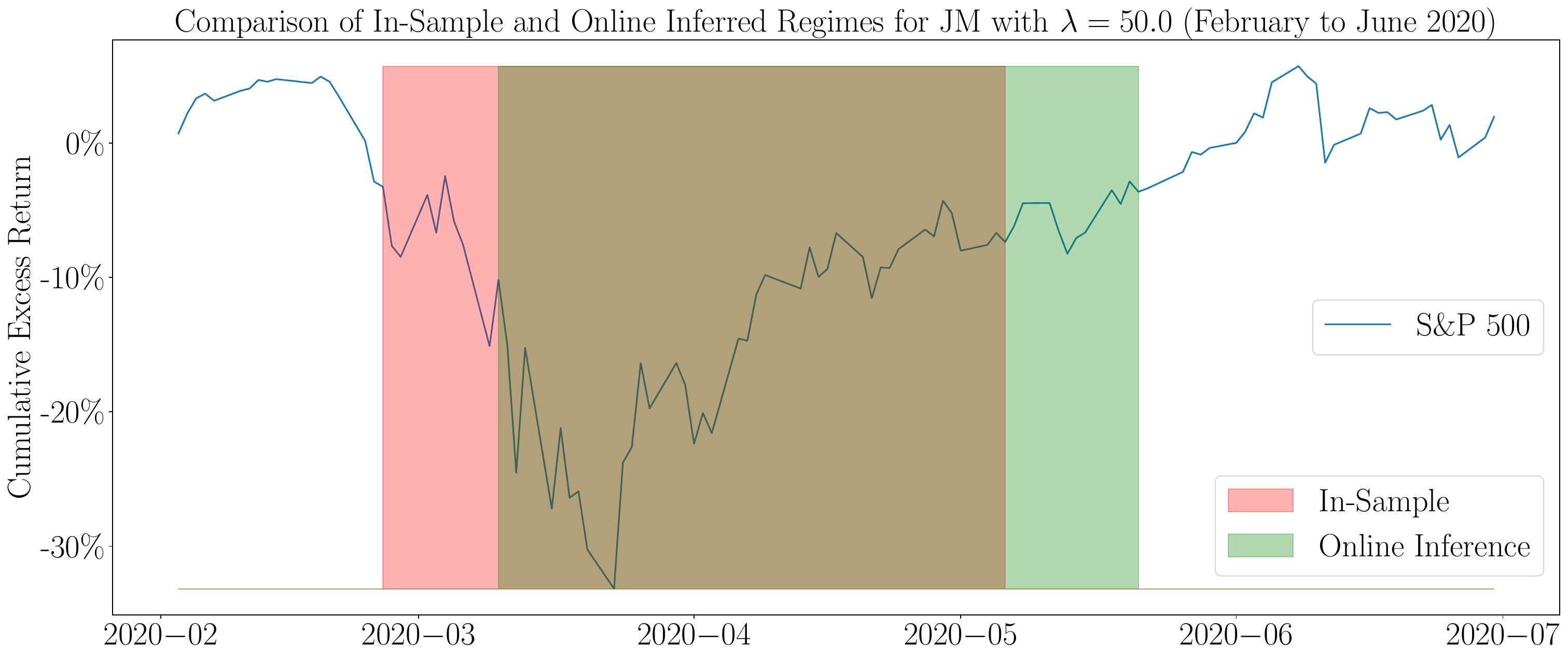}
    \caption{\small Comparison of regimes derived from in-sample training (in red) and those inferred online (in green) for the S\&P 500 index by JMs, with a fixed jump penalty of 50.0, during the COVID-19 crash period from February to June 2020.
    The in-sample regimes are based on a 3000-day training window ending at the end of 2020.}

    \label{fig:is oos comp}
\end{figure}

Before delving into the tactic for tuning the smoothing hyperparameters, we illustrate the latency of our online generated signal in indicating a regime shift after a \emph{turning point}, typically recognized only in hindsight, has already occurred.
Here, \emph{latency} refers to the model's timeliness in detecting a regime shift, distinct from the trading \emph{delay} which pertains to practical constraints.
Such latency is inherent in regime identification models, representing a tradeoff with accuracy.   %
For example, immediately after the onset of a bear market (identified ex-post), the model requires several days of negative returns to confirm a regime shift, in order to avoid frequent false alarms that compromise identification accuracy.
To demonstrate this, Figure \ref{fig:is oos comp} compares the regimes derived from in-sample training (shown in red) with those inferred online (shown in green) for the \mbox{S\&P 500} index by JMs ($\lambda$ fixed at 50.0) during the COVID-19 market crash from February to June 2020. 
The in-sample identified regimes are based on a 3000-day training window ending at the end of 2020, using complete market data up to that point, while the online inferred regimes are generated daily using only the latest available data. 
We observe a latency of approximately half a month in detecting both the onset and conclusion of the market crash, which is representative in our study and aligns with typical results in similar studies. 
For instance, \citet{Nystrup2018JPM} report that ``the median [latency] in detecting regime changes is 25 (calendar) days.''
During this period, the US market rebounded swiftly due to the Federal Reserve’s quantitative easing. 
Although using the online inferred regimes with a one-day trading delay did not necessarily improve strategy returns -- owing to latency at the end of the crash and the consequent missed market rebounds -- it effectively prevents a significant drawdown of about 20\%, thereby fulfilling the primary goal of reducing downside risk.

\subsubsection{Optimal Jump Penalty Selection} \label{subsec:hyper tuning}

Unlike previous studies that select the jump penalty based on statistical criteria such as classification accuracy \citep{nystrup2020jump, Aydinhan2024} or information criteria \citep{Cortese2024GIC}, we directly optimize the 0/1 strategy's performance\footnote{
A one-way transaction cost of 10 basis points is applied in calculating strategy performance.
} 
to maximize the practical benefits for the associated financial application.
Specifically, we use a time-series cross-validation approach, updating the optimal jump penalty monthly.
At the beginning of each month during the out-of-sample testing period, for each candidate jump penalty, we generate the online inferred regime sequence $\lp{\hat s_t}$ over an 8-year lookback validation window\footnote{
Given the noted discrepancy between in-sample fitting and online inference, it is recommended for applications using online inferred regimes to base model selection on these online inferred regimes rather than those identified in-sample.
This recommendation diverges from practices in studies like \citet{nystrup2020jump, Aydinhan2024, Bosancic2024}, which optimize an in-sample metric for model selection in JMs.
}, following the process as described in the previous subsection. 
This sequence is applied to trading with a one-day delay, and we evaluate the validation performance of the 0/1 strategy using these signals, effectively simulating a live-sample scenario, for each jump penalty. 
We then select the value $\hat\lambda$ that yields the highest Sharpe ratio during this validation period and use this value for the following month to generate out-of-sample regime inference post hyperparameter tuning. 
The optimally selected $\hat\lambda$ is also subject to a one-day trading delay; \mbox{i.e.}, $\hat\lambda$ selected at the end of day $t$ is applied to day $t+2$. 
Since our data begin in 1970, with training windows spanning 12 years and validation windows 8 years, the out-of-sample testing period begins in 1990. 
We employ the same method to optimally select the smoothing hyperparameter $k$ for HMMs.

\section{Empirical Results} \label{sec:results}

Table \ref{tab:perf_2} presents a performance comparison between the buy-and-old strategy and the 0/1 strategy using online inferred regimes from both the HMM and JM, with their respective hyperparameters optimally selected. 
These strategies are applied individually to the S\&P 500, DAX and Nikkei 225 indices over the testing period from 1990 to 2023.
Throughout the study, we impose a conservative transaction cost of 10 basis points for each one-way trade.

\begin{table}[htbp]
    \centering
    \begin{tabular}{l*{9}{r}}
\toprule
&\multicolumn{3}{c}{S\&P 500} & \multicolumn{3}{c}{DAX} &\multicolumn{3}{c}{Nikkei 225} \\
    \cmidrule(lr){2-4} \cmidrule(lr){5-7} \cmidrule(lr){8-10} 
 & B \& H & HMM & JM & B \& H & HMM & JM & B \& H & HMM & JM \\
\midrule
Return & 10.2\% & 8.5\% & 11.2\% & 6.8\% & 6.4\% & 8.6\% & 0.8\% & 2.5\% & 4.7\% \\
Volatility & 18.2\% & 11.3\% & 13.1\% & 22.1\% & 14.0\% & 16.4\% & 23.4\% & 16.0\% & 17.1\% \\
Sharpe & 0.48 & 0.54 & 0.68 & 0.30 & 0.35 & 0.44 & 0.12 & 0.19 & 0.31 \\
MDD & -55.2\% & -28.9\% & -26.6\% & -72.7\% & -40.5\% & -39.4\% & -79.1\% & -48.6\% & -45.3\% \\
Calmar & 0.16 & 0.21 & 0.33 & 0.09 & 0.12 & 0.18 & 0.04 & 0.06 & 0.12 \\
$\text{ES}_{0.05}$ & -2.7\% & -1.8\% & -2.0\% & -3.3\% & -2.2\% & -2.5\% & -3.4\% & -2.5\% & -2.6\% \\
Turnover & 0\% & 141\% & 44\% & 0\% & 246\% & 170\% & 0\% & 290\% & 72\% \\
Leverage & 100\% & 72\% & 80\% & 100\% & 73\% & 84\% & 100\% & 68\% & 75\% \\
\bottomrule
\end{tabular}

    \caption{\small Performance comparison of the buy-and-hold (``B \& H'') and the 0/1 strategies using online inferred regimes from the HMM and JM on the S\&P 500, DAX and Nikkei 225 indices from 1990 to 2023.
    Hyperparameters in HMMs and JMs are tuned as described in Section \ref{subsec:hyper tuning}.
    Each row represents an annualized performance metric: compound annual growth rate (``Return'', including the risk-free rate), volatility, Sharpe ratio (average excess return over volatility), maximum drawdown (``MDD''), Calmar ratio (average excess return over MDD), expected shortfall at 5\% level, portfolio turnover and leverage.
    A one-way transaction cost of 10 basis points is applied.
    }

    \label{tab:perf_2}
\end{table}

From the performance metrics presented, we first observe that the 0/1 strategy effectively improves several risk measures, including volatility, maximum drawdown (MDD), and expected shortfall at 5\% level.
Employing HMMs already significantly reduces MDD with the 0/1 strategy, while JMs further enhance this reduction. 
The expected shortfall also improves by approximately 1\%.  
These two metrics are often considered worst-case risk measures, and their improvement underscores the utility of the 0/1 strategy for downside risk protection, crucial for the long-term success of portfolio managers.

Regarding return and volatility measures, their values are related to the reduced leverage inherent in the 0/1 strategy.
Assuming, for simplicity, no transaction costs and an independent and identically distributed (\mbox{i.i.d.}) process for the return series, if \mbox{$100\times(1-l)$\%} of days are randomly selected to avoid market exposure in order to achieve a reduced leverage of $l<1$, then the return would be decreased by a factor of $l$, while risk would be decreased by a factor of $\sqrt l$, leading to a Sharpe ratio decreased by a factor of $\sqrt l$.
For example, for the S\&P 500 index, the JM-guided 0/1 strategy achieves a leverage of 80\%, leading to a baseline volatility of $18.2\%\times\sqrt{0.8}=16.3\%$, while our strategy further reduces volatility to 13.1\%.
Despite the reduced leverage, using JM improves the strategy's returns compared to the buy-and-hold strategy, enhancing the compound annual growth rate (CAGR) by 1\%, 1.8\% and 3.9\% for the three indices, respectively, and outperforming the HMM strategy, which achieves a CAGR on par with the index itself. 
Consequently, our JM strategy achieves higher risk-adjusted return metrics, including Sharpe and Calmar ratios, than both buy-and-holding the index and the HMM-guided strategy.
This illustrates the advantage of shifting to safer assets during accurately identified persistent bear markets.   %

The benefits of enhanced persistence offered by JMs are further substantiated by the significantly reduced turnover of the JM strategy, which compares  favorably with the HMM strategy.
The turnover of the JM-guided 0/1 strategy applied to the S\&P 500 is as low as 44\%, meaning that on average, the portfolio manager buys and sells 44\% of total allocation (a combined 88\% trading) each year  -- a relatively mild figure despite the large portfolio rebalancing that occurs with each regime shift. 
This demonstrates the practical applicability of our strategy.

\begin{figure}[htbp]
    \centering
    \includegraphics[width=\textwidth]{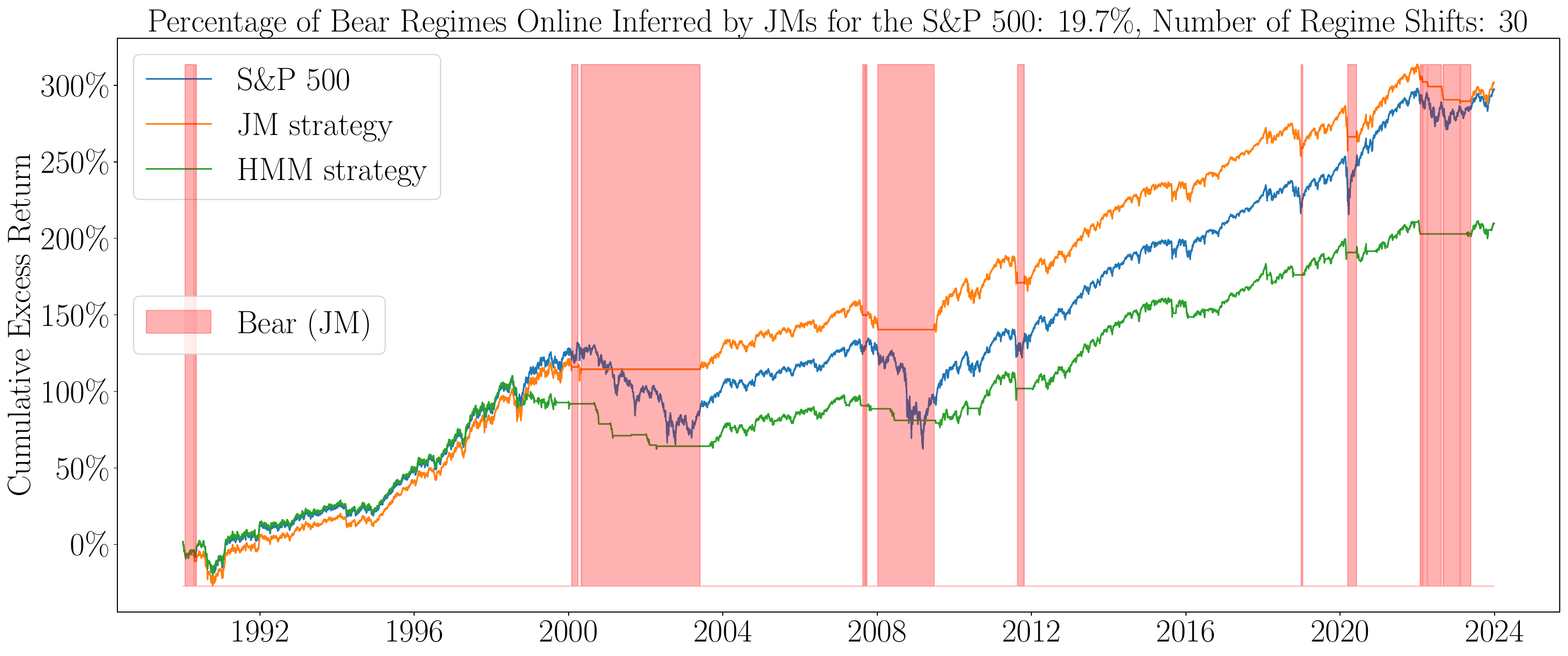}

    \includegraphics[width=\textwidth]{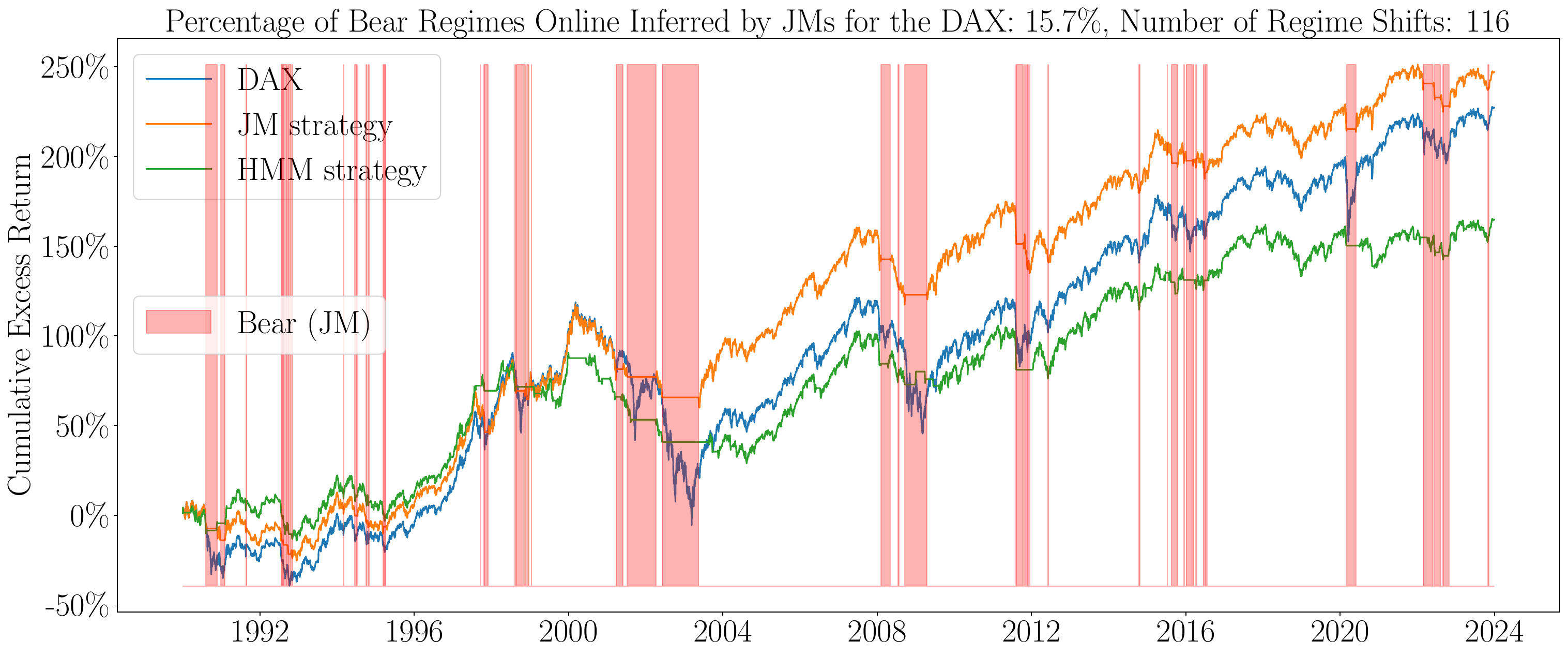}

    \includegraphics[width=\textwidth]{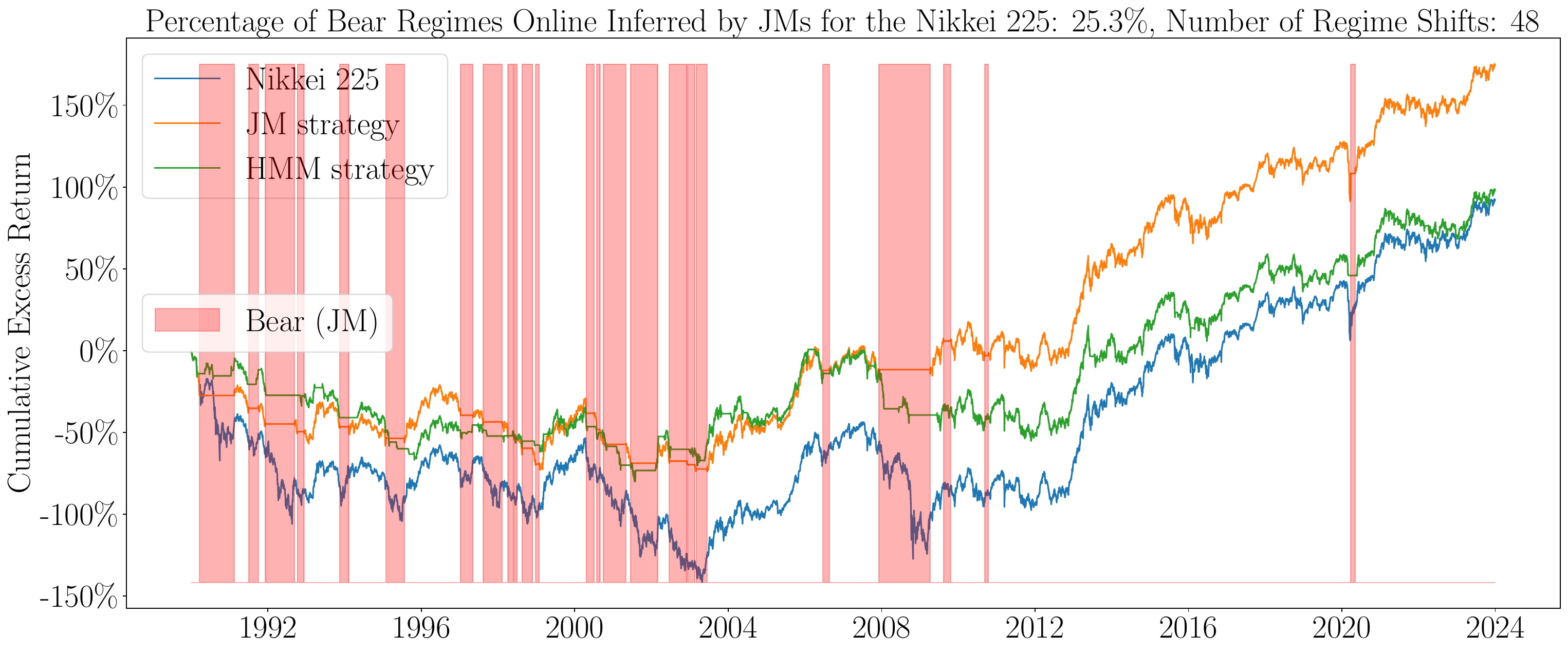}
    \caption{\small JM-inferred bear regimes and cumulative excess return curves for three strategies: the buy-and-hold (blue), the 0/1 strategy using JM-inferred regimes (yellow), and the 0/1 strategy using HMM-inferred regimes (green), for the S\&P 500, DAX and Nikkei 225 indices from 1990 to 2023.
    The red shading represents bear regimes online inferred by JMs (shifted forward by 2 days), when the JM-guided 0/1 strategy is fully invested in the risk-free asset, leading to a flat yellow curve.
    A one-way transaction cost of 10 basis points is applied.
    }

    \label{fig:regime plots}
\end{figure}

Figure \ref{fig:regime plots} provides a visualization of the bear regimes online inferred by JMs (shown in red shading) for the three indices, alongside the curves of cumulative excess returns from the three strategies compared in Table \ref{tab:perf_2}, spanning from 1990 to 2023. 
The bear regimes are shifted forward by two days to account for the trading delay, so that the red shaded areas represent periods when the JM-guided 0/1 strategy allocates 100\% to the risk-free asset, leading to a flat yellow curve. 
These regimes generally align with significant market downturns, such as the dot-com bubble in the early 2000s, the 2008 financial crisis, the 2020 COVID-19 crash, and the 2022 market decline driven by inflation and geopolitical tensions. 
Additionally, other region-specific bear regimes are captured, such as those experienced in Japan during the 1990s following the asset bubble burst.
The yellow curve representing the JM strategy often outperforms the others, exhibiting milder drawdowns and providing more robust protection against adverse market movements.
It consistently ends up with the highest returns across all indices.

Nonetheless, the identified regimes show some latency at the beginnings and endings of market crashes and might sometimes misinterpret the oscillations during prolonged turbulent periods, attributable to sharp positive and negative fluctuations. 
A potential enhancement could involve including more descriptive features that either detect trending or oscillatory patterns in the return series, or reflect broader macroeconomic conditions, to better inform JMs.

\begin{figure}[htbp]
    \centering
    \includegraphics[width=\textwidth]{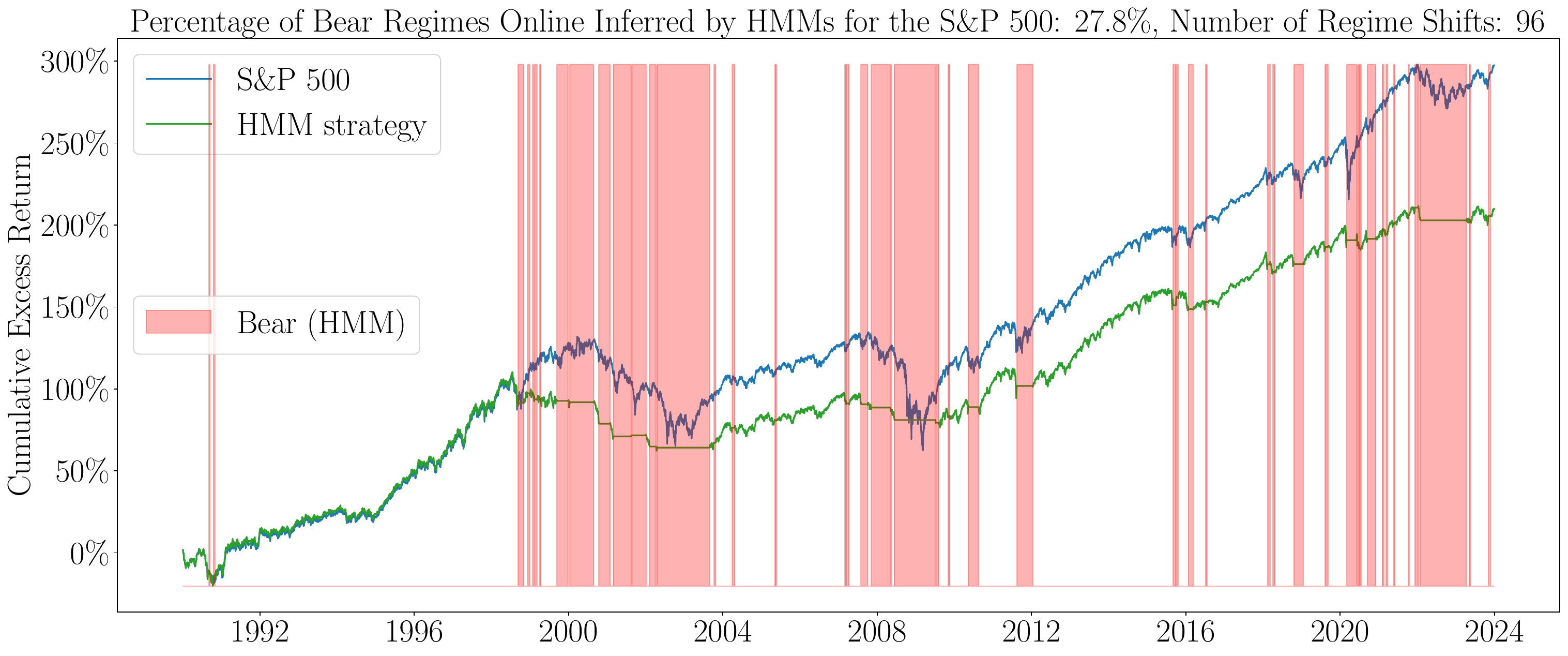}
    \caption{\small HMM-inferred bear regimes (red shading) and cumulative excess return curves for two strategies: the buy-and-hold (blue), the 0/1 strategy using HMM-inferred regimes (green), for the S\&P 500 index from 1990 to 2023.
    For additional details, refer to the caption of Figure \ref{fig:regime plots}.
    }  
    \label{fig:regime SPX HMM}
\end{figure}

Figure \ref{fig:regime SPX HMM} shows the HMM-inferred bear regimes for the S\&P 500 index, alongside its strategy's cumulative excess return curve from 1990 to 2023. 
In stark contrast to the JM-inferred regimes shown in Figure \ref{fig:regime plots}, the HMM-inferred regimes display numerous short-lived regimes that are unintuitive and difficult to trade, frequently resulting in the underperformance of the HMM strategy compared to the buy-and-hold strategy, attributable to both increased trading costs and compromised identification accuracy.
These short-lived regimes arise primarily from the HMM's sensitivity to daily market noise, often corresponding to brief periods of market oscillations that are challenging to time effectively.
Additionally, the HMM identifies several periods before 2000, during the pre-dot-com bubble, characterized by both high volatility and high returns. 
The HMM's emphasis on volatility limits its ability to exploit the upside potential during such market anomalies.

\subsection{Robustness Test: Trading Delays}  \label{subsec:sensitivity}

We conclude with a basic robustness test concerning the trading delay. 
Our previous discussions assume a one-day delay, meaning that the identified regime for day $t$ is applied to trading at the end of the following day $t+1$.
However, operational constraints might extend this delay, for instance, awaiting management approval. 
A potentially longer delay necessitates even greater persistence from our regime switching signals to maintain effectiveness.

Table \ref{tab:perf_all_delays} displays the performance of the 0/1 strategy under trading delays of 1, 5 (one week), and 10 (two weeks) days; effectively, the signal from the end of day $t$ is applied to days $t+2$,  $t+6$, and  $t+11$, respectively. 
For brevity, we present only (risk-adjusted) return metrics here. 
It is worth noting that we align the validation period with the out-of-sample testing period by assuming the same delay, ensuring that the cross-validation method automatically selects the appropriate level of smoothing for different trading delays. 
Naturally, assuming a one-day delay during validation but experiencing longer delays in a live-sample testing setting would degrade performance monotonically as the delay increases.
The optimally selected smoothing hyperparameter is also subject to the corresponding trading delay.

\begin{table}[htbp]
    \centering
    \begin{tabular}{l*{7}{r}}
\toprule
& \textbf{S\&P 500} & \multicolumn{3}{c}{HMM} & \multicolumn{3}{c}{JM}  \\
   \cmidrule(lr){2-2} \cmidrule(lr){3-5} \cmidrule(lr){6-8} 
 & B \& H & $\text{delay} = 1$  & 5& 10 & 1 & 5 & 10  \\
\midrule
Return & 10.2\% & 8.5\% & 8.6\% & 8.3\% & 11.2\% & 11.4\% & 11.7\% \\
Sharpe & 0.48 & 0.54 & 0.55 & 0.51 & 0.68 & 0.71 & 0.70 \\
Calmar & 0.16 & 0.21 & 0.28 & 0.25 & 0.33 & 0.39 & 0.28 \\
\midrule

& \textbf{DAX} & \multicolumn{3}{c}{HMM} & \multicolumn{3}{c}{JM}  \\
   \cmidrule(lr){2-2} \cmidrule(lr){3-5} \cmidrule(lr){6-8} 
 & B \& H & $\text{delay} = 1$  & 5& 10 & 1 & 5 & 10  \\
\midrule
Return & 6.8\% & 6.4\% & 5.1\% & 3.1\% & 8.6\% & 7.5\% & 5.9\% \\
Sharpe & 0.30 & 0.35 & 0.25 & 0.12 & 0.44 & 0.38 & 0.29 \\
Calmar & 0.09 & 0.12 & 0.09 & 0.04 & 0.18 & 0.13 & 0.11 \\

\midrule

& \textbf{Nikkei 225} & \multicolumn{3}{c}{HMM} & \multicolumn{3}{c}{JM}  \\
   \cmidrule(lr){2-2} \cmidrule(lr){3-5} \cmidrule(lr){6-8} 
 & B \& H & $\text{delay} = 1$  & 5& 10 & 1 & 5 & 10  \\
\midrule
Return & 0.8\% & 2.5\% & 1.2\% & 0.5\% & 4.7\% & 4.0\% & 3.4\% \\
Sharpe & 0.12 & 0.19 & 0.11 & 0.07 & 0.31 & 0.27 & 0.24 \\
Calmar & 0.04 & 0.06 & 0.03 & 0.02 & 0.12 & 0.09 & 0.07 \\

\bottomrule
\end{tabular}

    \caption{\small Performance comparison of the buy-and-hold (``B \& H'') and the 0/1 strategies using online inferred regimes from the HMM and JM, under trading delays of 1, 5, and 10 days for the S\&P 500, DAX and \mbox{Nikkei 225} indices (from top to bottom) form 1990 to 2023.
    A corresponding trading delay is also applied in the cross-validation when selecting the optimal smoothing hyperparameter.
    A one-way transaction cost of 10 basis points is applied.
    For details on the performance metrics, refer to the caption of Table \ref{tab:perf_2}.
}

    \label{tab:perf_all_delays}
\end{table}

Despite accommodation during validation, strategy performance generally declines with longer trading delays, though at varying rates.
For the S\&P 500, the decay is very slow for both JMs and HMMs, likely due to the high persistence of its regimes. 
In the case of the DAX and Nikkei 225 indices, the persistence of JMs becomes evident; the HMM strategy underperforms the market with a 5-day delay, whereas the JM strategy maintains a Sharpe ratio better than or comparable to the index even at the longest two-week delay. 
Thus, the JM strategy exhibits enhanced robustness to trading delays.

\section{Conclusion} \label{sec:conclusion}

In this article, we have presented a regime-switching investment strategy utilizing the statistical jump model (JM) to mitigate downside risk by timely shifting to safer assets in response to anticipated unfavorable market conditions. 
Our JM employs a set of features including risk and return measures derived from the return series, with the optimal jump penalty determined through a time-series cross-validation method that emphasizes the financial implications of identification accuracy. 
The flexibility and persistence of JMs provide practical advantages over traditional Markov-switching models in financial applications, such as reducing unnecessary portfolio rebalancing and accommodating a broader range of features.

Empirically, we have tested the strategy on major equity indices from the US, Germany, and Japan from 1990 to 2023, accounting for transaction costs and trading delays. 
The results indicate that the JM-guided strategy consistently outperforms both the hidden Markov model-guided strategy and the buy-and-hold strategy in reducing volatility and maximum drawdowns and enhancing risk-adjusted returns. 
Specifically, the JM-guided strategy improves annualized returns by approximately 1\% to 4\% over the buy-and-hold strategy across different regions.
Furthermore, JMs’ inherent persistence lends enhanced robustness against trading delays.
These empirical findings validate the effectiveness of JMs in real-world scenarios.

There are several promising directions for further research.
A more comprehensive feature set, together with a robust feature selection procedure, could refine the model’s accuracy and adaptability. 
It would be of interest to investigate how the influential features dynamically change over time.
Another opportunity is to extend the 0/1 strategy to a broader range of asset classes, including portfolios designed with specific characteristics such as long-short factor portfolios.
Additionally, further developments in JM methodologies, such as integrating more flexible distance measures beyond the standard $\ell_2$ distance, or employing distance measures in a latent space, hold potential for enhancing strategy performance.

{
\small

\bibliographystyle{apalike}
\bibliography{lit_bulla}

}

\end{document}